\documentclass[a4paper,10pt]{scrartcl}
\usepackage[utf8]{inputenc}
\usepackage{xcolor}
\usepackage{achemso}
\setkeys{acs}{usetitle = true, doi=true} 

\usepackage[version=4]{mhchem}
\usepackage{graphicx}
\usepackage[super]{nth}
\usepackage{placeins}
\usepackage{xspace}
\usepackage{mathtools}
\usepackage{wrapfig} 
\usepackage{amsmath} 

\title{On the Predictive Power of\\ Chemical Concepts}
\author{Stephanie A. Grimmel and Markus Reiher$^\star$}
\date{\small Laboratory of Physical Chemistry, ETH Zurich, Vladimir-Prelog-Weg 2, 8093 Zurich, Switzerland\\[2ex]
$^\star$ Corresponding author: markus.reiher@phys.chem.ethz.ch}

\begin{document}

\maketitle

\begin{abstract}
Many chemical concepts can be well defined in the context of quantum chemical theories.
Examples are the electronegativity scale of Mulliken and Jaffe and the hard and soft acids and bases concept of Pearson.
The sound theoretical basis allows for a systematic definition of such concepts. However, while they are often used to describe and compare chemical processes in terms of reactivity, their predictive power remains unclear.
In this work, we elaborate on the predictive potential of chemical reactivity concepts, which can be crucial 
for autonomous reaction exploration protocols to guide them by first-principles heuristics that exploit these concepts. 
\end{abstract}


\section{Introduction}
\label{sec:introduction}
Understanding chemical processes often starts with structure elucidation of the relevant reactive species and their connection in terms of a reaction network.
Quantum chemical calculations allow us to detail the reaction mechanism by mapping it to a network of elementary reaction steps characterized by reactants, stable intermediates, and products that are connected by transition states. In the most basic formulation of transition state theory, it is necessary to identify the stationary points, \textit{i.e.}, local minima and first-order saddle points, on a potential energy surface to arrive at such a reaction network at full conformational resolution. The mathematical definition of the potential energy surface may be taken from the Born-Oppenheimer approximation which freezes the nuclear motion compared to that of the electrons and introduces the electronic energy.
The electronic energy depends on the positions of the nuclei frozen in space. It can serve as the hypersurface for the
description of reactive as well as diffusive processes.
Quantum chemistry has provided routine tools to calculate and explore such surfaces\cite{warr86, cram06, jens17}, up to the point where this is possible in an automated manner\cite{same16, dewy18, simm19} in order to
provide the full detail of a chemical process in terms of thousands of interconnected elementary steps.
With subsequent inclusion of nuclear motion to arrive at free-energy differences for the elementary reaction steps, we have all information, obtained in a first-principles way with no experimental information required, to study the kinetics in the network\cite{kee80, kee89, kee96, hoop06, cantera, glow12, prop17, shan18, prop19}.

While this procedure provides us with a quantitative description of chemical reactions in terms of relevant
species involved, barrier heights, and concentration fluxes, it does not necessarily allow us to describe and
understand what is actually going on.
In particular, common reactivity patterns need to be identified from the quantitative data that is available from experiment and from detailed calculations.
In order to accomplish this, chemical concepts have been developed, many of them introduced in a rather \textit{ad hoc} fashion to describe changes in electronic structure upon a chemical reaction.
Examples are the electronegativity concept as defined by Pauling\cite{paul32, paul80} and the hard and soft acids and bases (HSAB) concept by Pearson\cite{pear63, pear68, pear68a}.
In the past decades, it has become possible to root these concepts in quantum chemical theories.
While this is natural for electronic structure related concepts such as partial charge and bond order, it is also possible for electronegativity and HSAB.

These concepts allow us to relate and discuss data obtained for different molecules and reactive systems.
As such, they have a categorizing value that can hardly be overestimated.
However, their predictive power remains obscure.
Even though rigorous mathematical definitions for many concepts exist, most concepts are not unique in a sense that they can be introduced in different ways (\textit{e.g.}, there exist various ways to define atomic and hence partial charge as an atom is not rigorously defined in a molecule without \textit{ad hoc} assumptions --- even Bader's elegant atoms in molecules theory\cite{bade90} is difficult to develop consistently in a fundamental relativistic framework without additional assumptions\cite{cios01, ande11}; loosely speaking, due to the lack of second derivatives in the Dirac equation\cite{reih15}).
Moreover, the concepts are typically evaluated for the unreactive, isolated reactants, \textit{i.e.}, in their equilibrium structures.
It is not at all clear how they could possibly point the way toward the minimum-energy reaction path and hence to the transition state
and barrier height.
Therefore, we may anticipate that it will be difficult to define a predictive concept that offers a rationale or better a quantitative
relation between the (molecular and/or electronic) structure of reactants and reactivity 
measures such as activation barriers (that eventually relate to rate constants, and hence, also to equilibrium constants).

Reactivity predictions should allow one to answer questions such as these:
Will certain reactants react with one another under given reaction conditions? 
Which reaction paths and/or products are to be expected, \textit{i.e.}, \textit{how} do the reactants react with each other? 
From which direction is a given reactant attacked preferably?
We focus on the prediction of elementary steps as opposed to more complex multistep reactions to which 
our discussion can be generalized in a straightforward manner.

For planning chemical synthesis, it is important to know whether and which of these concepts can be reliably employed.
Whereas one might argue that traditional chemical synthesis planning has rather been based on chemical intuition that is nourished by the vast amount of experimental knowledge (\textit{e.g.}, by name reactions) --- even when exploited in a data-driven manner\cite{kowa12, kluc18, szym16, schw19a, schw19, ibmrxn, segl17, segl17a, segl18}, recent developments in automated reaction exploration schemes\cite{same16, dewy18, simm19} based on extensive quantum chemical calculations would benefit from knowledge about the predictive value of reactivity concepts.

We have put forward the idea of first-principles heuristics, \textit{i.e.}, the idea of exploiting features of the electronic wave function to cut the deadwood from the reaction space,\cite{berg15, grim19} which for bimolecular reactions alone formally grows with the number of pairs of atoms in both reactants.
This can only be fruitful if reactivity predictions based on concepts are reliable (at least to a certain degree).
Therefore, we elicit on what is known about the predictive power of chemical reactivity concepts in this work.

For a concept to have predictive power, it should enable us to judge reliably on the reactivity of a given set of reactants without any \textit{a priori} information about possible reaction products and paths. 
Note, however, that it can be equally important --- especially in automated mechanism exploration --- to determine what atoms or functional groups in a molecule are unreactive. 

In the following section, we review mathematical definitions of four chemical concepts.
As examples we select the electronegativity and the chemical hardness, because they belong to the core of chemical
concepts taught to chemistry students from high school education onwards -- however, typically, without discussing their 
quantum chemical foundations.
Furthermore, we consider the Fukui functions and the dual descriptor 
for the identification of nucleophilic and electrophilic sites of reactants.
We outline limitations of specific descriptors as well as overall constraints.
Finally, we  expand on how to gather further insights upon the predictive capabilities of chemical concepts.
We elaborate on why, besides the value of chemical concepts for describing and understanding chemical processes,
massively automated reaction exploration algorithms will be important to assess actual reactive behavior in
full depth, shedding also light on the application range of reactivity concepts.

\section{Chemical Concepts}
Electronegativity, hard and soft acids and bases, Fukui functions, and the dual descriptor,
all refer to the electronic structure evaluated for a frozen molecular structure, typically the equilibrium
structure of a reactant. Hence, we limit ourselves to the standard electronic-energy based discussion
for specific individual structures on the Born-Oppenheimer surface, typically those that mark local minima.
Accordingly, thermal and entropic effects are left aside so that one can expect
predictions for enthalpic but not for entropic contributions. 
Naturally, these can only be sensible if the elementary step under
consideration is governed by electronic effects. 
We note, however, that there exist generalizations of quantum theory as well as chemical concepts
towards thermodynamic ensembles (see, \textit{e.g.},
Refs.~\citenum{mira18,gazq19}).

\subsection{Electronegativity}
\label{sec:electronegativity}
Electronegativity is among the oldest and most widely used chemical concepts.
The term dates back to Berzelius' work in the early \nth{19} century\cite{berz11} and the concept even further.\cite{jens96} 
It denotes the tendency of atoms to attract electrons towards themselves.\cite{paul32}
Among the many quantitative definitions that have been put forward, the first one by Pauling\cite{paul32} remains
to be the most popular one, which is why we briefly consider it here. Pauling
proposed a thermochemical recipe driven by the availability of such data at the time.
His definition relates differences in the electronegativity $\chi$ 
of two elements $A$ and $B$ to the deviation of the dissociation energy of the heteroatomic diatomic molecule, $E_\text{d}(AB)$, from the average of the dissociation energies of the corresponding homoatomic molecules, $E_\text{d}(AA)$ and $E_\text{d}(BB)$:
\begin{equation}
    \left| \chi^P_\text{A} - \chi^P_\text{B} \right| := \sqrt{E_\text{d}(AB) - \frac{E_\text{d}(AA) + E_\text{d}(BB)}{2}} \cdot (\text{eV}^{-1/2})
    \label{eq:Pauling}
\end{equation}
With regard to the predictive capabilities of the electronegativity it should be pointed out that by rearranging Eq.~\eqref{eq:Pauling} one obtains an expression for the heteroatomic dissociation energy in terms of Pauling electronegativities $\chi^P$ and homoatomic dissociation energies.
Considering the reliability of a heteroatomic dissociation energy calculated in this manner already sheds doubt 
on how reliable the electronegativities can potentially be.
Hence, examples of poor predictions can be easily found, \textit{e.g.}, in the case of alkali fluorides and chlorides.\cite{prit55}  

Indeed, the definition of $\chi^P$ through Eq.~\eqref{eq:Pauling} has profound limitations, some of which are:
\begin{enumerate}
\item If the evaluation of bond energies relies on enthalpies of formation (as in Pauling's work\cite{paul32}), it depends on macroscopic environmental parameters such as temperature.
\item The definition marries electronic with nuclear dynamics because experimental data was taken for the parametrization. 
As a consequence, the data cannot be easily calculated with sufficient accuracy.
\item It is not obvious whether only data on diatomics may be used for the definition or whether bond breaking processes in any molecule can be considered. \textit{I.e.}, the definition ignores the fact that atoms in different bonding situations may not be comparable. 
In other words, one will be forced to find a categorizing scheme that groups atoms in similar binding situations, for example, 
by introducing different valence states (and we refrain here from a discussion of how that could possibly be 
accomplished in a rigorous way). 
\item Moreover, Pauling's definition only allows for the calculation of relative electronegativity values, which requires an
arbitrary definition of an absolute reference. In his original work, he chose 0.0 for hydrogen, which was later replaced  by 2.1 and today is typically set to 2.2.\cite{paul32, paul80, hugg53, allr61}
\item The parametrization results in an overdetermined system of equations. 
\item Depending on which element pairs are employed the resulting electronegativities may differ. 
\end{enumerate}
From this list, it becomes clear that Pauling's scheme is ambiguous for a couple of reasons.
Most of these issues are not present in the absolute electronegativity definition by Mulliken,\cite{mull34}
which may be rigorously traced back to quantum chemical quantities that can be routinely calculated on modern computer hardware.
The Mulliken electronegativity $\chi^M$ of some atom is the average of its ionization energy $I$ and electron affinity $A$,
\begin{equation}
    \chi^M := \frac{I + A}{2} ,
    \label{eq:mulliken}
\end{equation}
which is a natural definition in the sense that it connects the concept to the two quantities that measure the energy required or
liberated when an electron is removed or added to the system. 
In our setting here, we may identify $I$ and $A$ as the vertical electronic energy differences calculated between the 
system's ground state and the corresponding states at the same structure with one electron removed and added, respectively
(with proper sign conventions).
With this definition, the concept of electronegativity is no longer limited to atoms but can be generalized to entire molecules.

An, at first sight, different approach to define electronegativity is taken in the context of conceptual density functional theory (cDFT) \cite{parr78, parr95, geer08}.
For an extensive review of the foundations of cDFT see Ref.~\citenum{geer03} and for a recent perspective Ref.~\citenum{geer20}.
cDFT provides mathematical definitions of numerous chemical concepts (see also below). Its basis is the
standard procedure of physical modeling and that is to expand a quantity that depends on a set of parameters in terms of a Taylor 
series. In this case, it is the electronic
energy $E_{el}[N, v;\{{\mathbf R}_k\}]$ that depends on the number of electrons $N$ and the external potential $v$. 
The latter is typically given as the 
Coulomb potential of the underlying nuclear framework in Born-Oppenheimer theory, and hence, it also encodes the specific molecular
structure given by the set of nuclear Cartesian coordinates of all nuclei, $\{{\mathbf R}_k\}$. 

Although this consideration is typically done in the framework of conceptual DFT, we emphasize that such a Taylor expansion 
of the electronic energy may be written
for any electronic structure model. However, DFT may offer specific advantages over other models (\textit{e.g.}, as 
Koopmans' theorem becomes exact; however, see our discussion below).
The identification of reactivity concepts as responses of $E_{el}$ towards changes in $N$ and $v$ is in line
with the understanding that they vary upon interactions between reactants during chemical reactions.

In 1961 -- and hence even before cDFT evolved as a field -- Iczkowski and Margrave identified the electronegativity with the negative response of the energy towards changes in the number of electrons at constant external potential:\cite{iczk61}
\begin{equation}
    \chi := - \left(\frac{\partial E_{el}}{\partial N}\right)_v = - \mu
    \label{eq:cdft_electronegativity}
\end{equation}
Later on, this expression was recognized as the negative of the chemical potential $\mu$ by Parr.\cite{parr78}
The minus sign is introduced to have a high (positive) electronegativity flag a strong tendency to attract electrons, and hence, 
it should be associated with a decrease in energy upon an increase in $N$ (note that for sufficiently large molecular
systems, the electronic energy can be considered to decrease upon reduction of the system by one electron).
Despite appearing  mathematically concise, Eq.~\eqref{eq:cdft_electronegativity} suffers from the fact that the number of electrons $N$ is  
restricted to integer numbers when treating isolated atoms or molecules.
This issue is discussed in great detail elsewhere.\cite{parr80, lieb83, drei90, cher99,geer03, kvaa14, geer20}

In practical applications, Eq.~\eqref{eq:cdft_electronegativity} is most often evaluated within a finite difference approach.
The electronegativity is then assumed to be equal to the average of the left and right derivatives of Eq.~\eqref{eq:cdft_electronegativity}, $\chi^-$ and $\chi^+$:\cite{geer03}
\begin{align}
    \chi^- &\approx E_{el}(N-1) - E_{el}(N) = I \\
    \chi^+ &\approx E_{el}(N) - E_{el}(N+1) = A \\
    \chi &\approx \frac{\chi^- +  \chi^+}{2} \approx \frac{I + A}{2} = \chi^M
    \label{eq:cdft_electronegativity_fda}
\end{align}
Even though derived differently, this expression is equal to the Mulliken electronegativity $\chi^M$ given in Eq.~\eqref{eq:mulliken}.
Therefore, finite differences clearly link both approaches.

Rather than calculating the electronic energy of the system with different electron numbers (\textit{i.e.}, $N$, $N+1$, and $N-1$),
one may replace ionization energy and electron affinity 
by orbital energies of the highest occupied molecular orbital (HOMO) and the lowest unoccupied molecular orbital (LUMO), 
$\epsilon_\text{HOMO}$ and $\epsilon_\text{LUMO}$,
respectively, of the system with $N$ electrons
by virtue of Koopmans' theorem\cite{koop34},
to obtain 
\begin{equation}
    \chi^M \approx -\frac{\epsilon_\text{HOMO} + \epsilon_\text{LUMO}}{2}.
\end{equation}
Whereas Koopmans' theorem relies on error cancellation in the case of Hartree-Fock
calculations\cite{mull49} (owing to the neglected orbital relaxation upon oxidation vs. the missing electron
correlation in the independent particle model), it can be considered exact for ionization energies 
in exact Kohn-Sham DFT\cite{perd82, almb85, perd97a}. However, the LUMO Kohn-Sham energy cannot be exploited that 
easily.\cite{baer13, baer18}  It refers to an excited
Kohn-Sham electron rather than to an added Kohn-Sham electron\cite{baer13, van14} so that one might consider the  
HOMO energy of the $N+1$ Kohn-Sham system and reverse its sign to obtain the electron affinity of the $N$ Kohn-Sham system.\cite{amat20}

Mulliken pointed out that the electronegativity is not only an atom's ground state property, but instead depends on its valence state.\cite{mull34}
This notion was formalized by Hinze and Jaffe\cite{hinz62, hinz63, hinz63b, hinz63a} by introduction of orbital electronegativities.
The electronegativity $\chi_i$ of orbital $i$ with $N_i$ being the number of electrons in that orbital is
given by
\begin{equation}
    \chi_i := - \left(\frac{\partial E_{el}}{\partial N_i}\right)_v  .
\end{equation}
Dramatically different electronegativity values were found, for example, for different valence states of carbon.\cite{berg96}

Although there is a decent correlation between Mulliken and Pauling electronegativities,\cite{skin53, hinz62, brat88} 
the theoretical basis of Mulliken-Jaffe electronegativities $\chi^M$ has the salient feature that it
links directly to quantities of electronic structure theory,
in which any (fixed) molecular scaffold represents a point on the Born-Oppenheimer surface given by the electronic energy,
as opposed to the profound problems associated with Pauling's definition. 

Electronegativities of atoms are widely used to identify reactive centers within molecules. For example, the fact that tetrahedral carbon has a lower electronegativity than chlorine (8.07 vs. 10.05~eV according to the Mulliken-Jaffe values from Ref.~\citenum{berg96}) allows us to predict that a nucleophilic attack on chloromethane should preferably occur at the carbon center.
The question whether electronegativity can be used to predict reaction barriers in a quantitative manner was tackled, \textit{e.g.}, in the context of hydrogen abstraction reactions:
Nguyen \textit{et al.} established a correlation between the reaction barriers of the abstraction of a given hydrogen atom from propene and methane and the electronegativity of attacking radicals.\cite{nguy04}
A larger share of chemical space was investigated in a study of the relation between different cDFT properties and 216 electrophilicity values from the Mayr database\cite{mayr05, mayr08, mayr11, mayr15} by Lee \textit{et al.}\cite{lee20}
They observed only a very poor correlation ($R^2 = 0.539$) between Mayr's electrophilicities and electronegativities obtained with Eq.~\eqref{eq:cdft_electronegativity_fda}.
An improved, but still weak correlation ($R^2 = 0.737$) was found with the cDFT electrophilicity\cite{parr99, szen00} $\omega$, which is a joint property calculated from the electronegativity $\chi$ and the chemical hardness $\eta$:
\begin{equation}
\omega :=\frac{\chi^2}{2\eta}
\label{eq:electrophilicity}
\end{equation}
The concept of chemical hardness will be considered in the following section.
The poor correlation coefficients may be taken as a clear indication that kinetic predictions based on Eqs.~\eqref{eq:cdft_electronegativity_fda} and~\eqref{eq:electrophilicity} are unreliable.
However, it should be pointed out that with Mayr's electrophilicities being obtained from experimental data, deviations may not be exclusively attributed to shortcomings of the concepts' definitions: Measuring inaccuracies and the protocol for the calculation of electrophilicities from experimental rate constants (see Ref.~\citenum{mayr08}) as well as limitations of the applied electronic structure method can also contribute to the discrepancies. This  issue is considered in some more detail in Section~\ref{sec:automated}.

\subsection{Hard and Soft Acids and Bases}
Based on experimental observations Pearson formulated the hard and soft acids and bases (HSAB) principle as follows: ``Hard acids prefer to bind to hard bases and soft acids prefer to bind to soft bases.''\cite{pear63, pear68, pear68a}
This means, that with \ce{A} denoting an acid, \ce{B} a base, and the subscripts $\text{h}$ and  $\text{s}$ hard and soft, respectively, the exchange reaction
\begin{equation*}
    \ce{A\textsubscript{h} B\textsubscript{s} + A\textsubscript{s} B\textsubscript{h} <=> A\textsubscript{h} B\textsubscript{h} + A\textsubscript{s} B\textsubscript{s}}
\end{equation*}
should be exergonic.
In Pearson's early work, soft acids and bases were defined to be those that are easily polarizable, while hard ones are hard to polarize.\cite{pear63} 
It is rather surprising that such a phenomenological concept that was proposed as an attempt to categorize a wealth of
experimental observations can actually be based on rigorous mathematical definitions in a quantum chemical framework.
cDFT provided such a framework, in which hardness $\eta$ appears as the resistance of the chemical potential $\mu$ 
towards changes in the number of electrons $N$. Accordingly, a derivative term in the aforementioned Taylor expansion of
the electronic energy is interpreted as the hardness:\cite{parr83}
\begin{equation}
    \tilde\eta := \frac{1}{2}\left(\frac{\partial^2 E_{el}}{\partial N^2} \right)_v = \frac{1}{2}\left(\frac{\partial \mu}{\partial N}\right)_v
    \label{eq:hardness_0}
\end{equation}
However, the prefactor 1/2 is frequently omitted:\cite{geer03, pear05}
\begin{equation}
    \eta := \left(\frac{\partial^2 E_{el}}{\partial N^2} \right)_v = \left(\frac{\partial \mu}{\partial N}\right)_v
    \label{eq:hardness}
\end{equation}
This means this prefactor emerging in front of any second-derivative term of a Taylor series expansion  
is to be understood as included in the definition, implying a scaling of the hardness. In the following, we continue to apply the scaled
hardness.

In analogy to electronegativity, the chemical hardness is often calculated within a finite difference approach,
which eventually reads as 
\begin{equation}
    \eta \approx I-A .
\end{equation}
The application of  Koopmans' theorem then yields
\begin{equation}
    \eta \approx \epsilon_\text{LUMO} - \epsilon_\text{HOMO}.
\end{equation}
The global softness, $S$, is identified as the inverse of the hardness,
\begin{equation}
    S := \frac{1}{\eta}.
\end{equation}
A prototypical example of Pearson's HSAB principle is the reaction between \ce{LiI} and \ce{CsF}:\cite{pear97}
\begin{equation*}
    \ce{LiI(g) + CsF(g) <=> LiF(g) + CsI(g)}
\end{equation*}
With \ce{I-} being softer than \ce{F-} and \ce{Cs+} softer than \ce{Li+}, the HSAB principle correctly predicts that the conversion should be exothermic. This is especially notable considering that a prediction based on Pauling electronegativities and a rearranged form of Eq.~\eqref{eq:Pauling} fails for this example.\cite{pear97}

Nevertheless, the HSAB principle should by no means be applied blindly. Most importantly, as Pearson himself pointed out,\cite{pear95} it only constitutes one amongst many effects determining the interaction between two reactants.
Considering the terms in the Taylor series expansion of the electronic energy,
this means that an exact correspondence between chemical reactivity and predictions based on the HSAB principle can only be expected if all other derivatives are negligible -- a condition that is hardly ever met in  chemical reactions.
For example, it is known that the HSAB principle tends to be dominated by the preference of strong acids to recombine with strong bases.\cite{card13}
Applications of the HSAB principle to ambident reactants were heavily criticized due to notable failures occurring even for prototypical systems.\cite{mayr11, bett20}

Although its now established roots in quantum chemistry allow for explicit calculations, the HSAB
principle should not be mistaken as a universal tool for reactivity predictions.
It may be regarded as one measure for reactivity in the context of all such concepts evaluated for a reactant.

\subsection{Fukui Indices and the Dual Descriptor}
To obtain information about \textit{where} a reactant is attacked, local reactivity descriptors are required.
One prominent example is the Fukui function $f(\mathbf{r})$ with the Cartesian coordinate $\mathbf{r}$ denoting some position in space.
It was introduced by Parr and Yang as a density-functional generalization of Fukui's Frontier Molecular Orbital (FMO) theory.\cite{parr84}

The Fukui function is the mixed second derivative of the electronic energy with respect to $v$ and $N$ at a given position $\mathbf{r}$.
It can be represented either as the response of the chemical potential to a change in the external potential at position $\mathbf{r}$ or as the response of the electron density $\rho(\mathbf{r})$ towards a change in the number of electrons, \textit{i.e.},
\begin{equation}
    f(\mathbf{r})  
:= \left(\frac{\delta}{\delta v(\mathbf{r})}\frac{\partial E_{el}}{\partial N}\right) 
=  \left(\frac{\delta \mu}{\delta v (\mathbf{r})}\right)
= \left(\frac{\partial}{\partial N}\frac{\delta E_{el}}{\delta v (\mathbf{r})}\right) 
= \left(\frac{\partial\rho(\mathbf{r})}{\partial N}\right).
\end{equation}
Regions where $f(\mathbf{r})$ is large are supposed to be those from which reactive attacks are the most favorable.\cite{parr84}
Due to the derivative discontinuities with respect to $N$, left and right derivatives are not equal, and hence, 
two different quantities are typically reported, $f^-(\mathbf{r})$ and $f^+(\mathbf{r})$, which can be
approximated by finite differences:
\begin{align}
f^-(\mathbf{r}) &:= \left(\frac{\partial\rho(\mathbf{r})}{\partial N}\right)_{v(\mathbf{r})}^- 
    \approx \rho_{N}(\mathbf{r}) - \rho_{N-1}(\mathbf{r}) 
\label{eq:fukui_minus}\\
f^+(\mathbf{r}) &:= \left(\frac{\partial\rho(\mathbf{r})}{\partial N}\right)_{v(\mathbf{r})}^+ 
    \approx \rho_{N+1}(\mathbf{r}) - \rho_{N}(\mathbf{r}) 
\label{eq:fukui_plus}
\end{align}
Hence, $f^-(\mathbf{r})$ describes the response of the system towards a decrease of $N$, \textit{i.e.}, towards electrophilic attacks and $f^+(\mathbf{r})$ 
the response towards an increase of $N$, \textit{i.e.}, the reactivity with respect to nucleophilic attacks.
To allow for an easier interpretation, the Fukui functions are often not analyzed in full spatial resolution, but instead in a condensed-to-atom representation.
Staying in the finite difference approximation,
these Fukui indices $f^\pm_k$ can be obtained for any atom $k$ from the atomic charges $q_k$ 
and are given in atomic units as
\begin{align}
    f^-_{k} &\approx {q_{k,N-1} -  q_{k,N}}\\
    f^+_{k} &\approx {q_{k,N}-  q_{k,N+1}}
\end{align}
However, atomic charges are yet another concept without a unique definition. Numerous mathematical definitions exist 
because of the difficulty to define spatial boundaries of an atom in a quantum system such as a molecule (see above).
Therefore, the choice among these can significantly affect the values of the resulting Fukui indices.\cite{bult07} 
Hirshfeld charges\cite{hirs77} are a commonly recommended choice\cite{prof02}.

If frozen orbitals are assumed, $f^-(\mathbf{r})$ and $f^+(\mathbf{r})$ can be expressed in terms of the removed or incoming orbital
upon change of the number of electrons, 
\begin{align}
    f^-(\mathbf{r}) &\approx \rho_\text{HOMO}(\mathbf{r})\\
    f^+(\mathbf{r}) &\approx \rho_\text{LUMO}(\mathbf{r}),
\end{align}
\textit{i.e.}, in terms of HOMO and LUMO densities, $\rho_\text{HOMO}(\mathbf{r})$ and $\rho_\text{LUMO}(\mathbf{r})$, respectively, calculated as the absolute square of the respective orbital,
which manifests the connection to FMO theory.
Naturally, this interpretation is prone to fail for quasi-degenerate frontier orbitals. For example, it cannot reliably predict the regioselectivity of electrophilic aromatic substitutions.\cite{dewa89, brow13}
In orbital-weighted Fukui functions this behavior is cured by the inclusion of further orbitals --- however, at the cost of introducing an empirically motivated weighting factor.\cite{pino17}

Instead of considering $f^-(\mathbf{r})$ and $f^+(\mathbf{r})$ separately to identify nucleophilic and electrophilic sites, Morell \textit{et al.} suggested the use of the dual descriptor $f^{(2)}(\mathbf{r})$ defined as the derivative of the chemical hardness with respect to the external potential or, equivalently, the response of the Fukui function towards changes in the number of electrons:\cite{more05, more06}
\begin{equation}
    \label{eq:dualdescriptor}
    f^{(2)}(\mathbf{r}) := \left(\frac{\delta \eta}{\delta v (\mathbf{r})}\right)_N = \left(\frac{\partial f (\mathbf{r})}{\partial N}\right)_{v(\mathbf{r})}
\end{equation}
Within the finite difference approximation, $f^{(2)}(\mathbf{r})$ can be written as the difference between the electrophilic and nucleophilic Fukui functions,
\begin{equation}
    \label{eq:dualdescriptor_fda}
    f^{(2)}(\mathbf{r}) \approx f^+(\mathbf{r}) - f^-(\mathbf{r}).
\end{equation}
$f^{(2)}(\mathbf{r})$ takes values between $-1$ and $1$ with negative and positive values indicating susceptibility towards electrophilic and nucleophilic attacks, respectively.\cite{more05} The dual descriptor was reported to be more robust than Fukui functions, especially in the frozen-orbital approximation.\cite{mart15a}

\begin{figure}[h]
    \centering
    \includegraphics[width=0.3\textwidth]{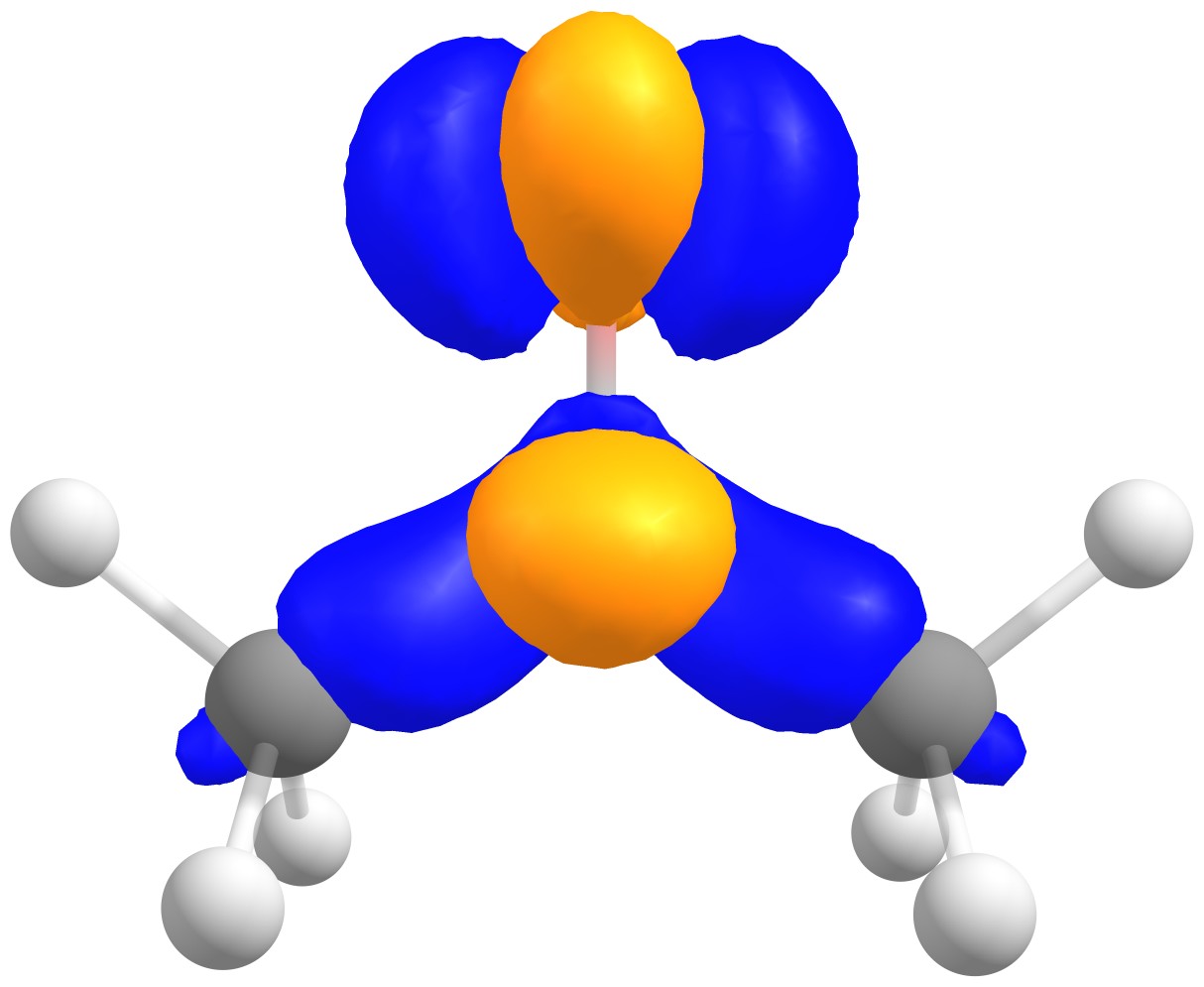}
    \caption{$f^{(2)}(\mathbf{r}) = \pm0.01\;\text{a}^{-3}_0$ isosurface for propanone. Blue encodes negative values, orange positive ones.}
    \label{fig:dualdescriptor}
\end{figure}

As a demonstration for the identification of reactive sites through the inspection of $f^{(2)}(\mathbf{r})$,
we consider propanone in the finite-difference approximation (see Fig.~\ref{fig:dualdescriptor}). 
We obtained the underlying electronic densities from a  DFT/PBE\cite{perd96, perd97}/def2-SVP\cite{weig05,weig06} calculation with the program \textsc{Orca}\cite{nees12}. The condensed-to-atom values given in Table~\ref{tab:dualdescriptor} were determined from Hirshfeld charges\cite{hirs77}. The fact that they turned out to be negative at the oxygen atom and positive at the adjacent carbon atom 
agrees with the notion that these are the primary attack sites for electrophilic and nucleophilic attacks, respectively.

\begin{table}[h]
    \centering
    \caption{Condensed dual descriptor indices for propanone.}
    \begin{tabular}{@{} *5l @{}}
        \hline
        \hline
        Atom & $f^{(2)}_\text{k}$ \\
        \hline
        O\textsubscript{CO} & $-$0.1 \\
        C\textsubscript{CO} & 0.1 \\
        C\textsubscript{$\alpha$} & 0.0 \\
        H & 0.0 \\
        \hline
        \hline
    \end{tabular}
    \label{tab:dualdescriptor}
\end{table}

Notably, the dual descriptor allowed for a back-on-the-envelope discussion of the Woodward-Hoffmann
rules\cite{ayer07a, geer12} refuting the common belief that one requires an orbital model to predict pericyclic reactions.

Unfortunately, the applicability of Fukui indices and the dual descriptor for predicting nucleophilicity and electrophilicity is limited to orbital-controlled soft--soft interactions.\cite{meli04} The most reactive sites for hard--hard interactions can be better predicted using atomic charges or with descriptors extracted from the molecular electrostatic potential.\cite{meli04, stuy20a} For example, Fukui functions are known to poorly predict protonation sites.\cite{meli04} Another level of complexity arises due to the fact that for many chemical reactions neither charge control nor orbital effects are highly dominating, but both have to be considered.\cite{ande07a, ande07, stuy20a}

\section{Limitations of Reactivity Prediction by Local Concepts}

As we outlined for the examples above, the predictive power of a single chemical concept is very limited.
Following the Taylor series expansion of the electronic energy, which explicitly states that various partial derivatives, now understood as different chemical concepts, are to be added up, it is not sufficient to limit oneself to one chemical concept when trying to understand the change in energy upon reaction, but several must be considered simultaneously.
They all need to be incorporated into a reactivity prediction protocol. 
The fact that machine-learned models based on several indicators clearly outperform schemes based on a single selected descriptor underlines this necessity.\cite{lee20a, hoff20a}

The key difficulty of all reactivity descriptors defined in terms of a Taylor series expansion of the electronic energy around a reference point is that this is always local information in the sense that it is valid for the unreactive and non-activated equilibrium structure of a reactant, \textit{i.e.}, for the reference point:
The local slope and curvature of the energy functional is supposed to correlate with the barrier height of eventual transition states.
As formulated in Klopman's non-crossing rule different reaction curves shall not cross between reactants and transition states.\cite{klop74}
In practice, this means that the behavior at the onset of reactions has to be decisive for the overall course of the reaction.
This is why Geerlings \textit{et al.} state that cDFT descriptors are expected to be unreliable for late transition states.\cite{geer20}
Making direct use of this statement for reactivity predictions, however, is impossible because transition states are obviously unknown prior to the analysis.

This issue may be considered more severe for all but intramolecular reactions due to the fact that to project towards reaction paths and energy barriers the interactions between all reactants have to be considered, which will disturb the information gathered at the reference point.
cDFT descriptors can be evaluated for associated reactive complexes and even along entire reaction paths,\cite{toro99} but the relevant complex geometry and especially the path are, obviously, not known \textit{a priori} in prediction tasks.
For a list of further approaches which incorporate the effect of a second reactant see Ref.~\citenum{geer20}.

In principle, it would be possible to evaluate derivatives of $E_{el}[N, v;\{{\mathbf R}_k\}]$ up to arbitrary order.
In practice, however, there are mathematical and technical difficulties associated with taking such derivatives.\cite{geer20}
Moreover, even if these derivatives were evaluated, there is no reason to believe that the radius of convergence of the series expansion is such that it encodes information up to the transition state structure.
In other words, the series expansion is not expected to converge for structures farther away from the reference point on the Born-Oppenheimer surface.

Due to these arguments, a perfect matching between reactivity descriptors, which are evaluated at equilibrium structures,
and activation barriers (and even less so for reaction energies) cannot be expected. They might, however, still be a valuable tool in order to \mbox{(de-)prioritize} more or less promising reaction guesses for further analyses. 
\textit{I.e.}, if one can tell (for instance, in an automated mechanism exploration algorithm), which sites are
very likely to be unreactive, then this knowledge accelerates the exploration as unreactive paths can be omitted
by the exploration protocol without the necessity to consider them explicitly.

\section{Reactivity Descriptors and Automated Reaction Space Exploration}
\label{sec:automated}
To verify the usefulness of a reactivity prediction scheme and gather profound knowledge about the associated uncertainties detailed statistical analyses have to be carried out.
Automated reaction space exploration schemes cannot only benefit from the exploitation of chemical concepts in the context of first-principles heuristics,\cite{berg15, grim19} they also offer the potential to help validating these.
For statistically rigorous analyses large and chemically diverse test sets have to be investigated. Automated algorithms allow to carry out and analyze vast numbers of quantum chemical calculations, inaccessible to manual work with quantum chemistry programs, and can therefore deliver the data required.

The compilation of such a test set is non-trivial: Typically researchers hand-pick reference reactions based on their prior knowledge or the literature or make use of databases such as the aforementioned Mayr database. However, this approach bears the risk of assembling biased test sets: Chemical concepts are integral parts of chemistry education.
That is why it appears reasonable to assume that reactions conforming to these are overrepresented in the literature and, hence, in the test sets.
This should not affect the comparison of different evaluation schemes for a given concept.
However, there might be a tendency to overrate the predictive power of chemical concepts as such in a circular-argument situation.
Automated reaction network exploration algorithms allow one to enumerate large numbers of different reaction coordinates --- including those that appear to be unpromising from a chemist's perspective.
Hence, there is hope that the bias in favor of known chemical concepts is less pronounced.

Furthermore, reaction databases typically only contain information on successful reactions, but no explicit information about unproductive reaction coordinates.
The possibility to sort out clearly unreactive reaction guesses is, however, one of the key tasks a useful reactivity predictor must fulfill, which is why this information should be included in a test set.
A brute-force enumeration of reaction paths by automated exploration algorithms will certainly also include unreactive scans.
Finally, the fact that the target quantity of the reactivity prediction (\textit{e.g.}, reaction barriers) is calculated with the same quantum mechanical method as the reactivity descriptor itself, opposed to experimental data, offers the possibility to study the reliability of the reactivity scheme separated from possible shortcomings of the underlying method compared to experiment.

So far, such extensive data has not been produced yet, but it can be anticipated that this will change\cite{same16, dewy18, simm19, unsl20}.

Once available, the machine learning evaluation of reactivity descriptors faced with explicit knowledge
of vast reaction networks will provide us with means to assess the extent to which we may prune quantum chemical reactivity explorations by first-principles
heuristics\cite{berg15, grim19} in order to dramatically accelerate the automated, but computationally costly exploration process.

\section{Conclusions}
In this work, we discussed why chemical concepts are not only valuable to explain, describe, and categorize reactions \textit{a posteriori},
but also bear hope for reactivity prediction.
Mathematical definitions allow for the quantification of well-known chemical concepts by means of quantum chemistry.
In particular, partial derivatives of the electronic energy with respect to the decisive parameters of a system (such as the number
of its electrons) allow us to clearly define reactivity concepts relying on electronic structure information. 

\begin{figure}[h]
    \centering
    \includegraphics[width=0.45\paperwidth]{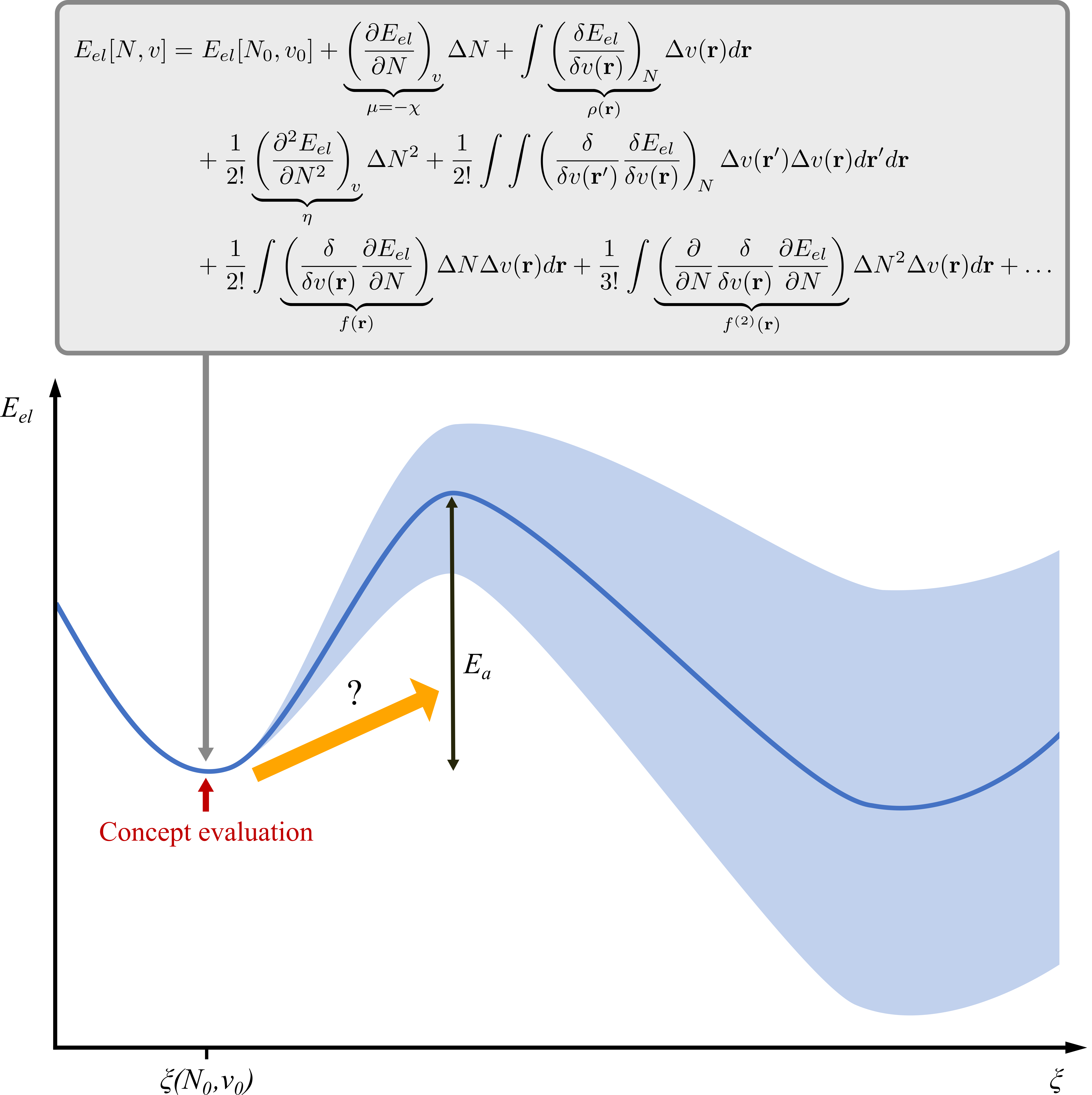}
    \caption{Concepts for stable reactant structures (red arrow) shall point to a reaction path $\xi$ with barrier~$E_a$ (orange arrow).
    Taylor expansion of the energy~$E_{el}$ around these structures (w.r.t. number of electrons~$N$ and external potential~$v$) delivers these concepts as partial derivatives.
    Reactivity prediction based on them will thus require all of them to be considered, but their local nature will lead to increasing uncertainty along the path (blue area).}
    \label{fig:reactioncurve}
\end{figure}

However, ambiguities may arise due to competing definitions of the same concept as well as different approximation and localization schemes.
An exact correspondence between chemical concepts and reactivity (measured, \textit{e.g.}, in terms of the activation barrier associated with that process) cannot be expected
because of the fact that the concepts are evaluated for a specific molecular structure, whereas barriers require the comparison of 
energies of two sets of structures, \textit{i.e.}, reactants and transition state structures (for an illustration of this problem see Fig.~\ref{fig:reactioncurve}).
The usefulness of chemical concepts critically depends on whether, despite these expected uncertainties, they still allow for discriminating between reactive and unreactive pathways with a high degree of confidence.

Especially automated quantum chemical reactivity exploration can effectively exploit this situation. If, due to the aforementioned uncertainties, we do not have clearly separable clusters of reactive and unreactive pairs of atoms, this will be a sign that our methodology cannot cut deadwood in
reaction space with sufficient precision. However, if we can define a trust margin that will allow us to include all 
of the potentially reactive sites, even at the cost of including some unreactive ones, then that will still represent a reliable basis for an efficient heuristic pruning scheme. 

Statistically rigorous analyses are required to quantify the reliability of reactivity prediction schemes
and to discover instances in which knowledge of reactant properties can, in fact, allow us to make useful predictions about
paths and barrier heights.
Automated reaction exploration algorithms are a suitable tool to provide the vast amounts of data required for such analyses.
They bear the promise to reduce the bias towards elementary steps that align with known concepts in the test sets and allow for efficient high-throughput analyses with modern statistical techniques.

Notorious machine learning techniques provide efficient algorithms to crawl through huge data sets produced in automated
mechanism explorations. They are a means to navigate reactivity data efficiently to highlight potentially important correlations
that then make the data accessible to our chemical understanding. As such, there is then no contradiction between the production and handling
of huge amounts of (calculated) reactivity data and the desire to understand the chemistry in terms of concepts that are
simple, rigorous, and effective. Concerns raised in the literature\cite{hoff20b} are likely
to be seen resolved in a fruitful interplay of numerical data and conceptual understanding. 

\section{Acknowledgments}
This work has been financially supported by the Swiss National Science Foundation (Project No. 200021\_182400).

\FloatBarrier
\addcontentsline{toc}{section}{Bibliography}
\FloatBarrier
\providecommand{\latin}[1]{#1}
\makeatletter
\providecommand{\doi}
  {\begingroup\let\do\@makeother\dospecials
  \catcode`\{=1 \catcode`\}=2 \doi@aux}
\providecommand{\doi@aux}[1]{\endgroup\texttt{#1}}
\makeatother
\providecommand*\mcitethebibliography{\thebibliography}
\csname @ifundefined\endcsname{endmcitethebibliography}
  {\let\endmcitethebibliography\endthebibliography}{}

\end{document}